\documentclass[10pt]{iopart}

\usepackage{graphicx}
\usepackage{times}

\begin{document}

\title{DNA uptake into nuclei: Numerical and analytical results}

\author{
 Z\'en\'o Farkas\dag\ddag,
 Imre Der\'enyi\dag\
 and
 Tam\'as Vicsek\dag
}

\address{
 \dag\ Department of Biological Physics, E\"otv\"os University,
 P\'azm\'any P. Stny 1A, H-1117 Budapest, Hungary}
\address{
 \ddag\ Department of Theoretical Physics, Gerhard-Mercator University,
 D-47048 Duisburg, Germany
}

\ead{
 zeno@angel.elte.hu,
 derenyi@angel.elte.hu,
 vicsek@angel.elte.hu
}

\begin{abstract}
The dynamics of polymer translocation through a pore has been the
subject of recent theoretical and experimental works. We have
considered theoretical estimates and performed computer simulations to
understand the mechanism of DNA uptake into the cell nucleus, a
phenomenon experimentally investigated by attaching a small bead to the
free end of the double helix and pulling this bead with the help of an
optical trap. The experiments show that the uptake is monotonous and
slows down when the remaining DNA segment becomes very short. Numerical
and analytical studies of the entropic repulsion between the DNA
filament and the membrane wall suggest a new interpretation of the
experimental observations. Our results indicate that the repulsion
monotonically decreases as the uptake progresses. Thus, the DNA is
pulled in (i) either by a small force of unknown origin, and then the
slowing down can be interpreted only statistically; (ii) or by a strong
but slow ratchet mechanism, which would naturally explain the observed
monotonicity, but then the slowing down requires additional
explanations. Only further experiments can unambiguously distinguish
between these two mechanisms.

\end{abstract}


\submitto{Journal of Physics: Condensed Matter}

\maketitle

\newcommand\mb[1]{\mbox{#1}}
\newcommand\idx[1]{{\mb{\scriptsize #1}}}
\newcommand\s{\,\mb{s}}
\newcommand\kg{\,\mb{kg}}
\newcommand\meter{\,\mb{m}}
\newcommand\nm{\,\mb{nm}}
\newcommand\pN{\,\mb{pN}}
\newcommand\micron{\,\mu\mb{m}}
\newcommand\Kelvin{\,\mb{K}}

\section{Introduction}
\label{sec:Introduction}

In eukaryotic cells molecular transport between the nucleus and the
cytoplasm takes place through the nuclear pore complexes (NPCs)
\cite{elbaum01,MBC}.
Ions and small proteins can diffuse through these large channels, but
large proteins and protein-RNA complexes are transported actively via a
signal-mediated process. DNA, on the other hand, is rarely transported
under normal circumstances. However, this latter process might be
relevant during viral infection or genetic therapy. In a recent
elegant experiment the uptake of single DNA molecules into the nucleus
was studied by the group of M.~Elbaum
\cite{salman01}.
In the present paper we provide a computational and theoretical study to
understand and interpret the results of this experiment.

\section{Description of the experiment}

\begin{figure}
\centerline{\includegraphics[width=0.67\linewidth]{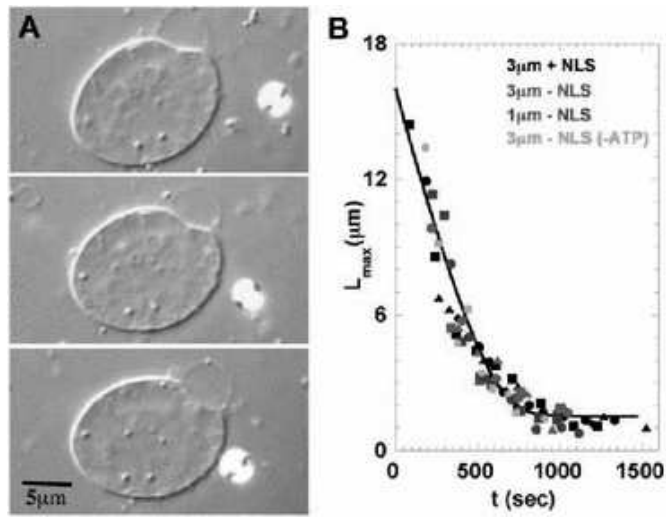}}
\caption{Kinetic measurement of DNA uptake. (A) The nucleus being
dragged after a $3 \micron$-diameter bead, linked by a DNA molecule.
The time intervals between recording the images are $532 \s$ and $302
\s$, respectively. Note the shortening of the extended length of the
DNA. (B) Measurements of the extended length vs.\ time.
The data was collected from many measurements. Since the initial moment
of association between the DNA and the nuclear pore was unknown, all
data was fitted to a common master curve.
(After \protect\cite{salman01}.)}
\label{fig:DNAExperiment}
\end{figure}

In the experiment double stranded bacteriophage
$\lambda$ DNA was transported into cell-free nuclei of {\em Xenopus
laevis} eggs. The DNA molecules were linked to polystyrene beads of $1$
or $3 \micron$. Beads lying near a nucleus were trapped by optical
tweezers and pulled away. In some cases dragging the bead pulled the
whole nucleus behind it at a distance (Fig.~\ref{fig:DNAExperiment}A).
After the bead was released, it started a biased Brownian motion due to
thermal fluctuations and its attachment to the DNA segment. Such
stretches were repeated on the same bead at two minutes intervals, and
the maximum extension decreased with each measurement
(Fig.~\ref{fig:DNAExperiment}B). This suggests that some mechanism
inside the nucleus pulled the DNA inward. The remaining extended
length vs.\ time curve has two main features: (i) it starts linearly,
and (ii) the length does not go to zero, instead, it converges to about
$1 \micron$. The uptake itself is very slow (compared to the time scale
of biochemical processes in living organisms), it takes several
hundred seconds for a $10 \micron$-long DNA.

\section{Theoretical estimates for the entropic forces}

To explain the convergence of the remaining length vs.\ time curve and
to give some estimations for the {\em uptake force} and the {\em
effective friction coefficient} of the pore, Salman et. al
\cite{salman01}
proposed a model in which the shrinking remaining segment of the DNA
applies an {\em increasing} resistance against the uptake.
They argue that because the entropic spring constant of a chain of
length $L$ is proportional to $L^{-1}$ and the average separation of
the two ends of the remaining DNA segment (caused by volume exclusion by
the bead and the nucleus) is proportional to $L^{1/2}$,
the resulting entropic repulsion diverges as their product, $L^{-1/2}$.

However, a detailed statistical analysis (described below) shows the
opposite: the entropic repulsion imposed by the shrinking DNA segment
{\em decreases} as $-L^{-1}$. This result makes it necessary to
reconsider the proposed kinetics of the uptake of DNA by the nucleus.

It has long been known that a polymer of length $L$, anchored at one of
its ends to a wall (with the other end being free), pulls on the wall
by the force
\begin{equation}
F_\idx{free ended} =
F_\idx{free ended}^\infty - (1-\gamma) \frac{k_\idx{B} T}{L}
\label{eq:Ffree}
\end{equation}  
at the anchoring point, on average
\cite{eisenriegler93,sung96,muthukumar99}.
$T$ denotes the absolute temperature, $k_\idx{B}$ is the Boltzmann
constant, and the length independent force term
$F_\idx{free ended}^\infty>0$ depends on the way of anchoring and on
other properties of the polymer. $\gamma\approx0.69$ for self-avoiding
and $\gamma=1/2$ for non-self-avoiding polymers. Due to force
balance the rest of the polymer pushes the wall by the same amount of
force.

Below we show that for non-self-avoiding polymers Eq.~(\ref{eq:Ffree})
can be obtained in a way much simpler than the original derivation,
using a simple {\em lattice chain} model. Then we develop further this
approach to take into account the role of the bead.

\begin{figure}
\centerline{\includegraphics[width=0.95\linewidth]{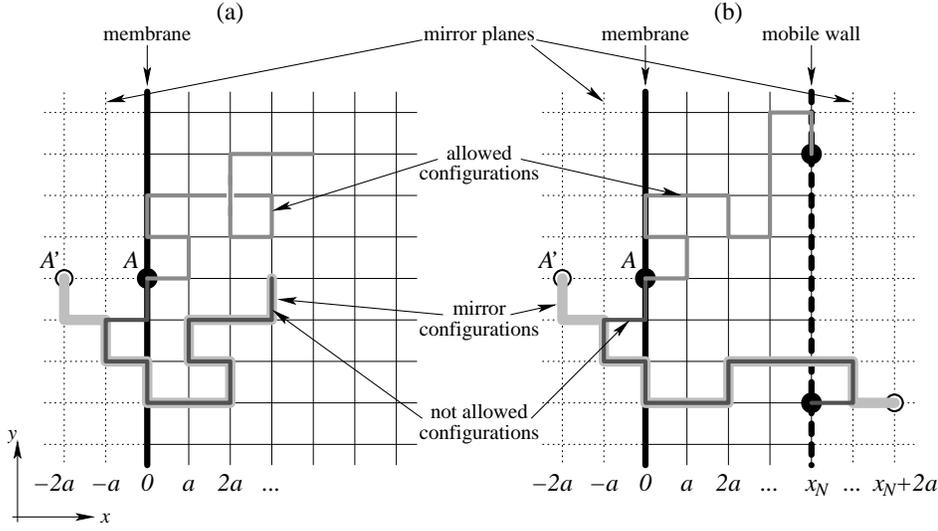}}
\caption{
Lattice chains of length $L=13a$, depicted in 2D. One end of the chains
is anchored to the membrane at point $A$, while the other end is either
(a) free, or (b) bound to a mobile wall. The figures indicate how to
construct the mirror configuration (light gray lines) of a
configuration not allowed by the constraints (dark gray lines).
}
\label{fig:lattice}
\end{figure}

If the {\em persistence length} of the polymer is $L_\idx{p}$, the lattice
constant of the equivalent three-dimensional lattice chain is $a=2
L_\idx{p}$, which is called the {\em Kuhn length}
\cite{doi86}.
Let us suppose that the wall is located at the $x=0$ plane, and the
anchoring point is at the origin $A=(0,0,0)$, as depicted in
Fig~\ref{fig:lattice}. The constraint that the chain is not allowed to
pierce through the wall is modeled by the condition that none of the
$x$ coordinates of the chain points can be negative: $0 \leq x_i$,
where $0 \leq i \leq N$ and $N=L/a$ is the number of Kuhn segments.
The partition function $\mathcal{Z}$ of this system is simply the sum of all
possible chain configurations that satisfy the above condition.

Let us choose an arbitrary point $B$ in the $0 \leq x$ half-space.
According to the so-called {\em reflection principle}, each
configuration that connects $A$ and $B$ but does not satisfy the
constraint, has one and only one ``mirror configuration'' that connects
$A'=(-2a,0,0)$ and $B$. Illustrated in Fig.~\ref{fig:lattice}a, the mirror
configuration is obtained by reflecting every chain point from $A$ and
to the first unsatisfied point about the $x=-a$ plane. Denoting the
number of unconstrained configurations that connect points $P$ and $Q$
by $\mathcal{N}(P\to Q)$, the partition function can be expressed as
\begin{equation}
\mathcal{Z}=\sum_{x_N \geq 0,y_N,z_N}
\Bigl\{
\mathcal{N}\left[A\to(x_N,y_N,z_N)\right] -
\mathcal{N}\left[A'\to(x_N,y_N,z_N)\right]
\Bigr\},
\label{eq:Zfree}
\end{equation}
where each configuration that is counted mistakenly in the first term
is canceled by its mirror configuration in the second term.
The partition function can further be written as
\begin{eqnarray}
\mathcal{Z}&=&\sum_{x_N \geq 0,y_N,z_N}
\Bigl\{
\mathcal{N}\left[A\to(x_N,y_N,z_N)\right] -
\mathcal{N}\left[A\to(x_N+2a,y_N,z_N)\right]
\Bigr\}
\nonumber\\
&=&\sum_{x_N\in\{0,a\},y_N,z_N}
\mathcal{N}\left[A\to(x_N,y_N,z_N)\right]
\\
&\approx& (2d)^N \, \mathcal{P}(0\leq x_N <2a) \approx (2d)^N \, 2a \, \rho(0) \; ,
\nonumber
\end{eqnarray}
where $d=3$ is the dimension of the space, $\mathcal{P}(0\leq x_N <2a)$
is the probability that the chain ends between the planes $x=0$ and
$x=2a$ in the continuum limit, and
$\rho(x)=(2\pi\sigma^2)^{-1/2}\exp[-x^2/(2\sigma^2)]$
is the probability density of the $x$ coordinate of the end position
with $\sigma^2=a^2 N/d$. The approximations are valid for large $N$.

Thus, the partition function can be expressed as
\begin{equation}
\mathcal{Z}\approx \pi^{-1/2} (2d)^N \left(\frac{2d}{N}\right)^{1/2}\, ,
\end{equation}
the free energy is
\begin{equation}
\mathcal{F}=-k_\idx{B} T\ln(\mathcal{Z}) \, ,
\end{equation}
and the force acting on the anchored end of the chain is
\begin{equation}
F_\idx{free ended}=-\frac{\partial \mathcal{F}}{\partial L}
=-\frac{1}{a}\frac{\partial \mathcal{F}}{\partial N}
\approx k_\idx{B} T\frac{\ln(2d)}{2 L_\idx{p}}
 -\frac{1}{2}\frac{k_\idx{B} T}{L} ,
\label{eq:Flatticefree}
\end{equation}
in agreement with Eq.~(\ref{eq:Ffree}).

In the experiments, however, the distal end of the polymer is not free,
but is bound to a polystyrene bead. Because the bead is large compared
to the remaining DNA coil when the uptake stalls, we can approximate it
with a mobile wall. The lattice chain calculation can be repeated even
for this system. The partition function, i.e., the number of allowed
configurations is now
\begin{eqnarray}
\mathcal{Z}&=&\sum_{x_N \geq 0,y_N,z_N}
\Bigl\{
\mathcal{N}\left[A\to(x_N,y_N,z_N)\right] -
\mathcal{N}\left[A'\to(x_N,y_N,z_N)\right]
\Bigr.-
\nonumber\\
&&-
\Bigl.
\mathcal{N}\left[A\to(x_N+2a,y_N,z_N)\right] +
\mathcal{N}\left[A'\to(x_N+2a,y_N,z_N)\right]
\Bigr\},
\end{eqnarray}
where the first two terms are from Eq.~(\ref{eq:Zfree}), the third
term -- based on the reflection principle again (see
Fig~\ref{fig:lattice}b) -- cancels all
configurations that go beyond the end point $(x_N,y_N,z_N)$, and the
forth term compensates for those configurations that have been
canceled twice (in both the second and the third terms). Replacing
$\mathcal{N}\left[A'\to(x_N,y_N,z_N)\right]$ by
$\mathcal{N}\left[A\to(x_N+2a,y_N,z_N)\right]$, and
$\mathcal{N}\left[A'\to(x_N+2a,y_N,z_N)\right]$ by
$\mathcal{N}\left[A\to(x_N+4a,y_N,z_N)\right]$, and making further
simplifications one arrives at
\begin{eqnarray}
\mathcal{Z}&=&
\sum_{x_N\in\{0,a\},y_N,z_N}
\mathcal{N}\left[A\to(x_N,y_N,z_N)\right]-
\nonumber\\
&&-
\sum_{x_N\in\{2a,3a\},y_N,z_N}
\mathcal{N}\left[A\to(x_N,y_N,z_N)\right]
\\
&\approx&
(2d)^N \, 2a \, [\rho(0)-\rho(2a)] \approx
\pi^{-1/2} (2d)^N \left(\frac{2d}{N}\right)^{3/2}\; ,
\nonumber
\end{eqnarray}
leading to the entropic force
\begin{equation}
F_\idx{mobile wall}
\approx k_\idx{B} T\frac{\ln(2d)}{2 L_\idx{p}}
-\frac{3}{2}\frac{k_\idx{B} T}{L} ,
\label{eq:Flatticewall}
\end{equation}
which indicates that a bead (or mobile wall) attached to the end of the
DNA renders the entropic resistance even smaller. This can be
understood intuitively by noticing that near the end of the uptake
process the geometric constraints drastically reduce the available
configuration space for the remaining segment, so it becomes more and
more favorable for the DNA to slide through to the other side of the
membrane.

Although the reflection principle can only be applied to
non-self-avoiding polymers, we expect a qualitatively similar behavior
for self-avoiding ones, too. In this latter case the reduction of the
configuration space should also decrease the entropic resistance.

\begin{figure}
\centerline{\includegraphics[height=0.67\linewidth,angle=270]{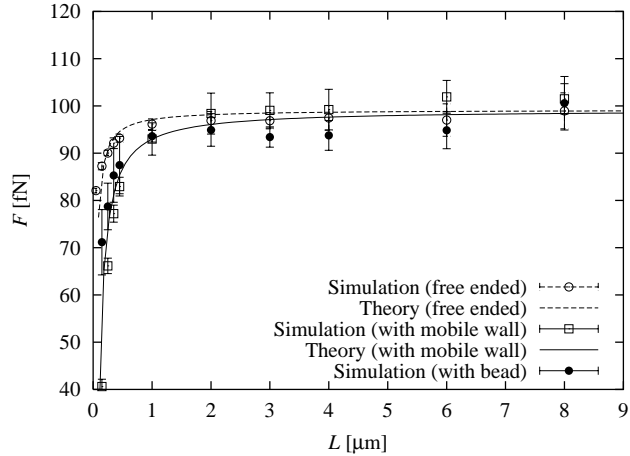}}
\caption{The force exerted by the membrane as a function of the
polymer length. The symbols are the results of numerical simulations
with $b = 50 \nm$, averaged over ten 2-second-long runs.
The curves are theoretical predictions (shifted upward to fit the data
set).
}
\label{fig:ForceEntr}
\end{figure}

To verify our theoretical estimations we have performed numerical
simulations. As Fig.~\ref{fig:ForceEntr} indicates, the numerical
results for the free ended filament and for the one with the mobile
wall are in good agreement with the relevant terms 
$-k_\idx{B}T/(2L)$ and $-3k_\idx{B}T/(2L)$ of the theoretical
predictions, respectively. The case with the polystyrene bead, just as
expected, is in between the above two extremes, closer to the one with
the mobile wall.

\section{Simulation technique}

\begin{figure}
\centerline{\includegraphics[width=0.4\linewidth]{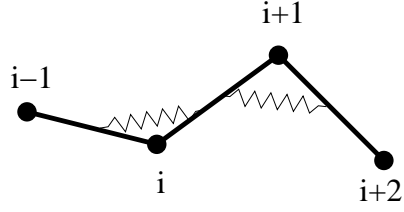}}
\caption{Discretized polymer: the beads are connected by the rigid
rods (bonds), and the bending elasticity is taken into account by
angular springs.}
\label{fig:WLC}
\end{figure}

We modeled the double stranded DNA as a three-dimensional ``bead-rod''
chain \cite{doi86}, and used molecular dynamics simulations to follow
its dynamics (see Fig.\ \ref{fig:WLC}). Note that the ``beads'' of the
bead-rod chain are artificial units and have nothing to do with the
polystyrene bead in the real experiment. Monte Carlo methods were ruled
out because they usually fail to reproduce the correct dynamics.

The overdamped equations of motion of the beads are
\begin{equation}
\Gamma \dot \mathbf{R}_j = \mathbf{F}_j^\idx{constr}
+ \mathbf{F}_j^\idx{bend} + \mathbf{F}_j^\idx{metr} 
+ \mathbf{F}_j^\idx{ext} + \xi_j,
\end{equation}
where $\Gamma$ is the friction coefficient and
$\mathbf{R}_j$ is the position of the $j$th bead ($1 \le j \le N+1$).
The bond vectors can be defined as $\mathbf{r}_j = \mathbf{R}_{j+1} -
\mathbf{R}_j$ (for $1 \le j \le N$).
The forces on the r.h.s.\ of the equations are as follows:
\begin{itemize}

\item $\mathbf{F}_j^\idx{constr}$ is the constraint force.
These forces are responsible for
keeping the length of the rods (or bonds) fixed: $|\mathbf{r}_j|=b$.
The constraint forces can be decomposed
into the bond tension forces: 
$\mathbf{F}_j^\idx{constr} = \mathbf{F}_j^\idx{tens} - \mathbf{F}_{j-1}^\idx{tens}$
with $\mathbf{F}_j^\idx{tens} = T_j \mathbf{r}_j$ and
$\mathbf{F}_0^\idx{tens} = \mathbf{F}_{N+1}^\idx{tens} = 0$
\cite{everaers99}.
The tensions $T_j$, represent $N$ new variables, however, we have
exactly this number of additional equations expressing that the bond
lengths are constant:
$\mathbf{r}_j \cdot \dot \mathbf{r}_j = 0$.

\item $\mathbf{F}_j^\idx{bend}$ is the bending force. The bending resistance
is represented by angular springs between the adjacent bonds.
The bending energy is 
$E^\idx{bend} = \frac{\kappa(b)}{2b} \sum_{i=1}^{N-1} \theta_i^2$, where
$\kappa(b)$ is the {\em bond length dependent} bending modulus,
$\theta_i = \arccos (\mathbf{r}_i \cdot \mathbf{r}_{i+1} / b^2)$
is the included angle between bonds $i$ and $i+1$. 
The bending force is thus $\mathbf{F}_j^\idx{bend} = -\nabla_j E^\idx{bend}$.
In the literature usually
the bending modulus of the continuous polymer, $\kappa_0$, is used to
calculate the bending energy even in a discretized polymer model.
However, we realized that for a discretized
bead-rod polymer the bending modulus should depend on the bond length
to give correct statistics. The detailed calculations will be
presented elsewhere, we show here only the result:
$\hat\kappa(\hat b) \approx 1 - \hat b / 6 + \hat b^2 (4/\pi^2-13/24)
+ \hat b^3 (1/6-2/\pi^2) + \hat b^{4} / 96$, where
$\hat b = b / L_\idx{p}$ is the dimensionless bond length
and $\hat\kappa(\hat b) = \kappa(b) / \kappa_0$
is the dimensionless bending modulus.
The persistence length $L_\idx{p}$ of the polymer is related to the bending
modulus through the expression $L_\idx{p} = \kappa_0 / (k_\idx{B} T)$.
A feature of this bond length dependent bending modulus is that
$\kappa (2 L_\idx{p}) = 0$, which reflects the known fact that
a continuous worm-like chain with persistence length $L_\idx{p}$ can be well
approximated by a freely jointed chain with bond length 
$b = 2 L_\idx{p}$ on the large scale.

\item $\mathbf{F}_j^\idx{metr}$ is the {\em metric} or {\em pseudo} force,
and is needed because the motion of a bead-rod chain in a thermal bath
is different from the limiting case of very stiff bonds
\cite{fixman74,fixman78,hinch94,hinch95,hinch96}.
The latter is
a better model for real biopolymers, and the metric force turns the rigid
rods into very stiff springs.
$\mathbf{F}_j^\idx{metr} = -k_\idx{B} T \nabla_j \ln 
\sqrt{|g|}$, where $g$ is the metric tensor of the hypersurface of the
constrained $N+1$ beads in the originally $3(N+1)$ dimensional
configuration space. The explicit form of this metric tensor for a
linear bead-rod chain can be found in Refs.
\cite{fixman74,hinch96}.

\item $\mathbf{F}_j^\idx{ext}$ is an optional external force acting on
bead $j$. In the application presented in this paper, the polystyrene bead
is attached to the end of the polymer by applying an attracting
external force on the last polymer bead, while this bead and the wall
of the nucleus exert
repelling forces on the polymer.  Depending on the situation, the
bond length $b$ may be too large to
represent the filament contour faithfully.
For example, in the simulation the DNA molecule 
goes through a $10 \nm$ wide pore embedded in a $100 \nm$ thick
membrane wall, while the bod length is $100 \nm$. The spatial
resolution of the bead-rod chain in this case would be very low
compared to the size of the pore. For this reason we introduced $M$
``virtual'' beads
along each rod, spaced uniformly.  Their sole role is that external
forces can act on them, and they appropriately redistribute these
forces to the original beads at both ends of the bond they are sitting on.
Hence their presence improve the spatial resolution, as far as
external forces are concerned. We used $M = 10$ in the simulations.

\item $\xi_j$ represents spatially and temporally uncorrelated 
thermal noise, obeying the fluctuation-dissipation relation
$\langle \xi_j(t) \xi_k(t')\rangle = 2 k_\idx{B} T \Gamma \delta_{jk} 
\delta(t-t')$.

\end{itemize}

In our model we neglected the excluded volume effect for the DNA filament
itself, although we accounted for the exclusion of the filament by the
polystyrene bead and the nuclear membrane wall.
Hydrodynamic interactions were also neglected.

In our simulations the persistence length was $L_\idx{p} = 50 \nm$
and the friction coefficient of the beads was
$\Gamma \approx \pi\eta b$, where the viscosity of the water is 
$\eta = 10^{-3} \kg \s^{-1} \meter^{-1}$ and the temperature is
$T = 300 \Kelvin$.
We used the second-order Runge-Kutta (or midpoint) method to solve
the system of first-order ordinary equations \cite{numrec}.
The bond length $b$ varied between $50 \nm$ and $100 \nm$.

One can rely on the results of a dynamical simulation only if it is
numerically stable. Therefore, it is important to be aware of the
physical processes that determine the largest possible time step.
In our case the diffusion of the polymer beads and the
relaxation of the angular springs pose an upper limit for the time
step. The typical distance a diffusive particle takes during
time $\Delta t$ is $\Delta x = \sqrt{2 D \Delta t}$, where the
diffusion constant is $D = k_\idx{B} T / \Gamma$. To ensure that the
typical change of the bond length is smaller than a small fraction
$\varepsilon_D b$ of the bond length, the time step cannot be larger than
$$
\Delta t_D = \frac{\tilde\Gamma \varepsilon^2_D}{2 k_\idx{B} T} b^3,
$$
where $\tilde\Gamma = \Gamma / b$ is the friction coefficient per unit
length. Here the constraint forces were not taken into
account. However, we expect that even with the constraint forces in
effect the numerical error is still proportional to the 
typical diffusive displacement of the beads in one time step. 

The relaxation time of the angular springs is \cite{farkasphd}
$$
\tau_{AS} = \frac{\tilde\Gamma}{6 \kappa(b)} b^4.
$$
Apart from the prefactor, this can be easily seen from dimensional
analysis. Thus, the time step cannot be larger than
$\Delta t_{AS} = \varepsilon_{AS} \tau_{AS}$ either, where
$\varepsilon_{AS} \ll 1$.
In our simulations we used $\varepsilon_D = \varepsilon_{AS} = 0.05$.
The maximum allowable time step is then the minimum of $\Delta t_D$
and $\Delta t_{AS}$.

\section{Discussion}

A DNA filament threading through a nuclear pore is pulled by two
opposing entropic forces from the two sides of the membrane. If the
media on the two sides are identical, the length independent force
terms cancel each other. Because the length dependent contributions
(being proportional to $k_\idx{B}T/L$) disappear quickly for large
$L$, there must exist some active (energy consuming) mechanism
responsible for the pulling of the DNA. Since the integral of the
length dependent term (from $L_\idx{p}$ to the total length of the DNA)
is in the order of only a few $k_\idx{B}T$, this force term is not even
strong enough to trap the DNA in a configuration where almost the
entire DNA is on one side of the membrane.

\subsection{Chemical composition difference}

The simplest way to implement an active pulling mechanism is to produce
a difference between the length independent terms of the entropic
forces on the two sides of the membrane. This can be achieved, e.g., via
a difference in the ionic strengths or the pH values of the two media.
Even the simple lattice chain model can account for this effect if the
chemical composition of the medium affect the persistence length
$L_\idx{p}$ of the DNA [see Eqs.~(\ref{eq:Flatticefree}) and
(\ref{eq:Flatticewall})].

\subsection{Ratchet mechanism}

Another possibility for the pulling is based on a ratchet mechanism
\cite{elbaumprivate}.
The ratchet mechanism as a means of active transport was first proposed
by the group of G. Oster
\cite{simon92,peskin93}.
The idea is that inside the nucleus some kind of molecule is able to
bind to the DNA with a rate constant $k_\idx{on}$ and
to detach from it with a rate constant $k_\idx{off}$.
When such a molecule is bound to the DNA at some location,
the DNA cannot slide back beyond this point.

Assuming that the possible binding sites for the molecule are
separated by a distance $b$, we can estimate the effective pulling
force produced by this ratchet mechanism. 
The grand canonical partition function per binding site is
$\mathcal{Z} = 1 + e^{-(E_b-\mu)/(k_\idx{B} T)}$,
where $E_b$ is the binding energy and $\mu$ is the
chemical potential for the molecule. The probability that the
site is unoccupied is $p_\idx{off} = 1 / \mathcal{Z}$, and the
probability that it is occupied is
$p_\idx{on} = e^{-(E_b-\mu)/(k_\idx{B} T)}/\mathcal{Z}$.
Thus, the partition function can be written as
$\mathcal{Z} = 1 + p_\idx{on}/p_\idx{off} = 1 + k_\idx{on}/k_\idx{off}$. 
The free energies of a binding site inside and outside the nucleus are
$\mathcal{F}_\idx{in} = -k_\idx{B} T \ln \mathcal{Z}$ and
$\mathcal{F}_\idx{out} = -k_\idx{B} T \ln 1 = 0$, respectively.
Finally, an effective pulling force can be defined as
\begin{equation}
F_\idx{ratchet} = -(\mathcal{F}_\idx{in} - \mathcal{F}_\idx{out}) / b =
\ln \left( 1 + k_\idx{on}/k_\idx{off} \right) k_\idx{B} T / b.
\end{equation}
Note that this is the force necessary to stop the pulling. The pulling
speed is proportional to this effective force only if the equilibration
time of the binding of the molecules $(k_\idx{on}+k_\idx{off})^{-1}$ is
shorter than the diffusion time of the DNA over the distance $b$.
Otherwise the pulling speed is smaller.

\subsection{Friction in the pore}

The friction of the DNA inside the nuclear pores can significantly
slow down the uptake process. Let us just
estimate how long it would take for an entire DNA to diffuse through
the pore if there was no friction inside. The diffusion can be
envisioned as the subsequent diffusion of $N=L_\idx{DNA}/a$ Kuhn
segments (where $a=2L_\idx{p}$ is the Kuhn length). The diffusion time
of a Kuhn segment can be approximated as
$\tau_\idx{K}=(1/2) a^2 (\pi\eta a)/(k_\idx{B} T)$,
thus the diffusion time of the entire DNA is about
$\tau=N^2\tau_\idx{K}=L_\idx{DNA}^2 L_\idx{p} (\pi\eta)/(k_\idx{B} T)$.
As expected for a polymer without self-exclusion, the diffusion time is
proportional to the square of its length
\cite{kardar01}.

The diffusion time $\tau$ for a $15 \micron$ long DNA is in the order
of 10 sec. Because the uptake occurs much slower, in the order of 1000
sec, the friction in the pore must be at least 100-fold greater than
the hydrodynamic friction:
$\Gamma_\idx{pore} > 100 L_\idx{p} \pi\eta \approx 10^{-8}$kg/s.
Such a strong friction can be explained by weak binding of the DNA to
the nuclear pore
\cite{lubensky99}.

\subsection{Uptake kinetics I: weak pulling}

The experiments show that the uptake slows down in the end and the
remaining length of the DNA converges to $L\idx{conv}\approx 1\micron$.
A constant pulling force $F$ (produced by either a chemical
composition difference or a ratchet mechanism) corresponds to a linear
potential $U(L)=FL$ with a reflecting boundary at $L=0$. In such a
potential the expectation value of $L$ at equilibrium is given by
$k_\idx{B} T/F$. Thus, the measured equilibrium distance
$L\idx{conv}$ is compatible with an $F=k_\idx{B} T/L\idx{conv} \approx
4$fN pulling force. From the average pulling speed, $v\approx 2\times
10^{-8}$m/s, one can also calculate the friction coefficient:
$\Gamma_\idx{pore} = F/v \approx 2\times 10^{-7}$kg/s, which satisfies
the condition given in the preceding paragraph.

The pulling force and the friction coefficient would be even smaller if
the length dependent term of the entropic force was also taken into
account in the calculation of the equilibrium distance.

\begin{figure}
\centerline{\includegraphics[width=0.95\linewidth]{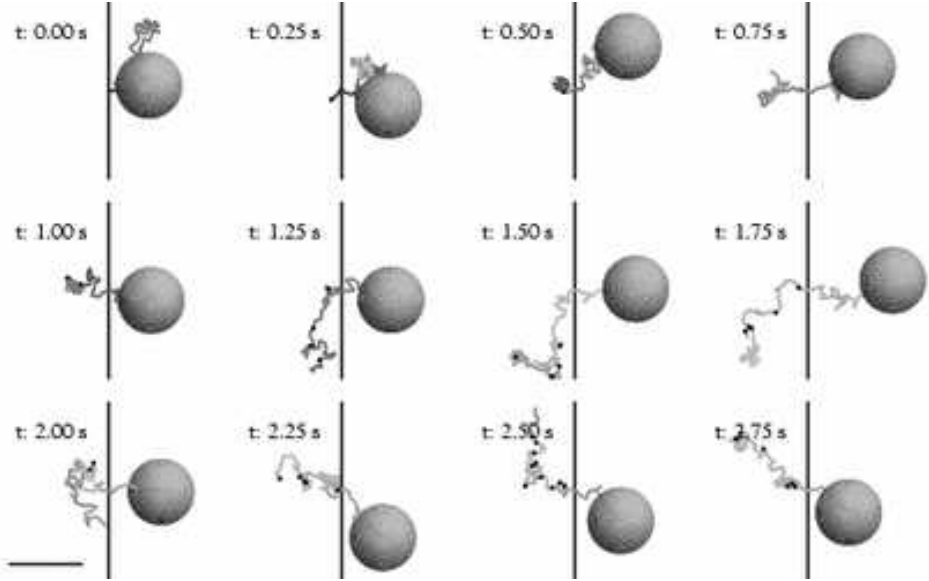}}
\caption{The uptake of a DNA filament. Its length is $8 \micron$,
the bond length is $b = 100 \nm$, and the radius of the polystyrene
bead is $R_\idx{b} = 500 \nm$. The black dots represent occupied binding sites.
The horizontal bar in the bottom left corner corresponds to $1 \micron$.
The gray scale indicates the position in the third dimension (depth).}
\label{fig:DNAUptakeSimulation}
\end{figure}

We studied the uptake process by the ratchet mechanism using numerical
simulations. We assumed
that the molecules could bind only to the beads of the bead-rod chain.
The simulations started from a configuration in which one end of the DNA
had already got through the pore, and the first bead was not allowed to slip
back through the pore.  To mimic the experiment, a large bead of
radius $R_\idx{b}$ was
attached to the other, outer end of the DNA.  The diameter of the pore
was $10 \nm$, the width of the wall was $100 \nm$, and the friction at
the pore was neglected.
Fig.~\ref{fig:DNAUptakeSimulation} shows a sequence of such a simulation.

Identifying the extended length of the remaining DNA by its arc length,
we plotted the extended length vs.\ time curve, averaged over 20 runs.
It resembles the experimental curve: it starts linearly and converges
to a constant value, about $1 \micron$, which coincides with the
experimental one. The time scale of the simulation is shorter than that
of the experiments, because we omitted the friction inside the pore, in
order to save computational time.

The negative side of this interpretation is that the uptake is stochastic.
Individual runs show back-slips at both the linear and the horizontal
part of the curve. The stationary state of the system is not an energy
minimum but a dynamic equilibrium between the fluctuations of the DNA
filament and the weak pulling of the transport mechanism. Since such
fluctuations are not apparent in the experiments, we examine another
possibility for the kinetics of the uptake.

\begin{figure}
\centerline{\includegraphics[height=0.7\linewidth,angle=270]{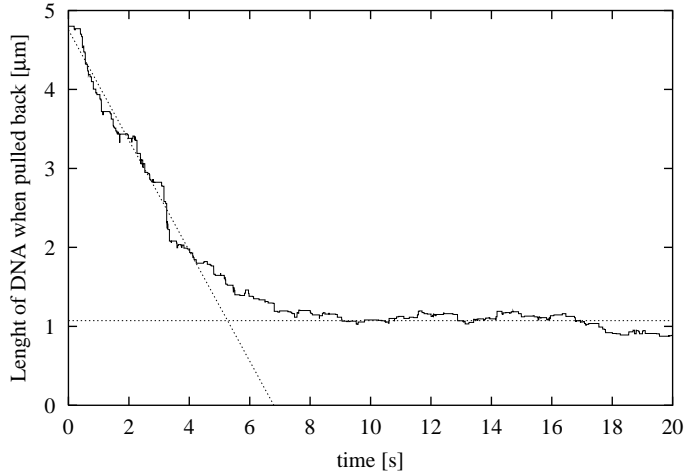}}
\caption{The remaining length of the DNA vs.\ time curve.
The parameters are $k_\idx{on} = 0.1 \s^{-1}$, $k_\idx{off} = 0.25 \s^{-1}$,
$b = 100 \nm$, $R_\idx{b} = 0.2 \micron$, and the total length of the DNA is
$4.8 \micron$. The curve is an average over 20 runs.}
\label{fig:DNAUptakeLen}
\end{figure}

\subsection{Uptake kinetics II: strong but slow ratcheting}

A monotonous uptake can be achieved either by a strong continuous
pulling in conjunction with a strong friction or by a strong but slow
ratchet mechanism. In such a ratchet mechanism the speed of the uptake
would be set by the small binding rate $k_\idx{on}$, however, the much
smaller unbinding rate $k_\idx{off}\ll k_\idx{on}$ would ensure that
unbinding is negligibly rare and, thus, the ratcheting is practically
irreversible. Note that the large $k_\idx{on}/k_\idx{off}$ ratio
corresponds to a large effective pulling force.

Due to the negligibly rare unbinding events the ratchet mechanism is
also supported by the fact that the uptake can be stopped but never
reversed in the experiments
\cite{salman01}
by applying pN forces on the polystyrene bead. A continuous
pulling would be incompatible with this observation.

The only phenomenon that cannot be explained by such a ratchet
mechanism is the convergence of the remaining length to a non-zero
value. There are, however, alternative explanations for this:
\begin{itemize}
\item the non-zero distance might be an artifact of the image
processing;
\item there might be unknown structures on the nuclear envelope that
prevent the polystyrene bead from getting in contact with the nucleus;
\item the pN forces used to extend the DNA might be sufficiently strong
to deform the membrane by hundreds of nm-s at the vicinity of the
pore
\cite{derenyi02,powers02};
\end{itemize}
or a combination of these.

\section{Conclusions}

Our numerical and analytical studies of the uptake of DNA filaments
into the nucleus suggest two possible mechanisms for the uptake
process. None of them can be ruled out on the basis of existing
experimental data. One of the mechanisms explains the convergence of
the remaining length of the DNA to a finite value, but has an
intrinsically stochastic nature (back-slips occur). This mechanism
alone cannot account for the stopping of the uptake under retarding
forces in the pN range. The other mechanism seems more plausible, it
predicts a monotonous uptake, but requires additional elements to
explain the convergence. Only further experiments can unambiguously
distinguish between these mechanisms, focusing on their differences.

\ack

We would like to acknowledge Collegium Budapest for its hospitality.
We also thank M. Elbaum for many useful discussions and
helpful comments. 
This research has been supported by the Hungarian National Research
Grant Foundation (OTKA T033104).


\end{document}